\documentclass[%
aip,
apl,%
amsmath,amssymb,
reprint,%
]{revtex4-2}

\usepackage{graphicx}
\usepackage{dcolumn}
\usepackage{bm}
\usepackage{sidecap}
\begin{document}
	
	\preprint{AIP/123-QED}

	\title{Huge magnetostriction in superconducting single-crystalline BaFe$_{1.908}$Ni$_{0.092}$As$_{2}$}
	
	\author{Minjie Zhang$ ^{1,2} $,Jiating Wu$ ^{1,2} $,Ke Shi$ ^{1,2} $,Langsheng Ling$ ^{1} $,Wei Tong$ ^{1} $,Chuanying Xi$ ^{1} $,Li Pi$ ^{1} $, J. Wosnitza$ ^{3,4} $,Huiqian Luo$ ^{5,6, a)} $,Zhaosheng Wang$ ^{1, a)} $}
	
	\affiliation
	{$ ^{1} $Anhui Key Laboratory of Condensed Matter Physics at Extreme Conditions, High Magnetic Field Laboratory, Hefei Institutes of Physical Science, Chinese Academy of Sciences, Hefei 230031, China  \\$ ^{2} $University of Science and Technology of China, Hefei 230026, China \\ $ ^{3} $Hochfeld-Magnetlabor Dresden (HLD-EMFL) and W{\"u}rzburg-Dresden Cluster of Excellence ct.qmat, Helmholtz-Zentrum Dresden-Rossendorf, 01328 Dresden, Germany \\$ ^{4} $Institut für Festk{\"o}rper- und Materialphysik, Technische Universit{\"a}t Dresden, 01062 Dresden, Germany \\  $^{5}$Beijing National Laboratory for Condensed Matter Physics, Institute of Physics, Chinese Academy of Sciences, Beijing 100190, China \\  $ ^{6} $Songshan Lake Materials Laboratory, Dongguan, Guangdong 523808, China \\ $ ^{a)} $Authors to whom correspondence should be addressed: hqluo@iphy.ac.cn and zswang@hmfl.ac.cn}

	\begin{abstract}
	    The performance of iron-based superconductors in high magnetic fields plays an important role for their practical application. 
	    In this work, we measured the magnetostriction and magnetization of BaFe$_{1.908}$Ni$_{0.092}$As$_{2}$ single crystals using pulsed magnetic fields up to 60 T and static magnetic fields up to 33 T, respectively. 
	    A huge longitudinal magnetostriction (of the order of 10$ ^{-4} $) was observed in the direction of the twin boundaries. 
		The magnetization measurements evidence a high critical-current density due to strong bulk pinning.  
		By using magnetization data with an exponential flux-pinning model, we can reproduce the magnetostriction curves qualitatively. 
		This result shows that the magnetostriction of BaFe$_{1.908}$Ni$_{0.092}$As$_{2}$ can be well explained by a flux-pinning-induced mechanism. 
	\end{abstract}
	
	\maketitle
	
	During the intense research on high-temperature superconductors, new compounds and synthetic routes have been tested in an effort to create materials with better properties for practical applications. 
	Since the discovery of iron-based superconductors (IBSs), numerous studies have been devoted to the understanding of the superconducting mechanism and their application potential. \cite{01,02,03}
	Among the existing IBSs, the 122-type compounds were extensively studied due to their unique properties, such as easy growth and relative stability. \cite{a,b,c}
	They are considered to be the most competitive candidates for practical high-field applications, because of their relatively high transition temperature $ T_{c} $, excellent transport properties with high critical currents $J_{c} $, very high upper critical field $ H_{c2} $ with relatively small anisotropy. \cite{04,05,06,07} 
	To date, breakthroughs have been made in the technical application of 122-type IBSs as bulk magnets, thin films, and wires. \cite{04} 
	In particular, the successful development of the world's first 100-m-class-length (Sr,K)Fe$ _{2} $As$ _{2} $ wire with an average $ J_{c} $ of 1.3 $ \times $ 10$ ^{4} $ A/cm$ ^{2} $ and the optimization of Ba$ _{0.6} $K$ _{0.4} $Fe$ _{2} $As$ _{2} $ tapes achieving $ J_{c} $ up to 1.5 $\times$ 10$^{5}$ A/cm$^{2}$ are of great significance. \cite{08,09}
	
	However, higher critical-current densities and higher trapped fields also implies greater pinning-induced strain in a superconductor.  
	In 1993, Ikuta $ et  $  al.\ discovered that Bi$ _{2} $Sr$ _{2} $CaCu$ _{2} $O$ _{8} $ single crystals deformed by a relative $ \Delta $$ L $/$ L $ compression exceeding 10$ ^{-4} $ in response to magnetic fields and proposed a flux-pinning-induced magnetostriction mechanism to explain the newfound giant deformation effect. \cite{10}
	Consequently, a number of groups studied typical questions of practical significance such as the location of maximum magnetostriction of high-temperature superconductors and the influence of sample shape on the magnitude and type of the deformation. \cite{11,12,13,14,16,17,18}
	For bulk superconductors, the magnetization process in large magnetic fields causes pinning-induced strain that leads to internal fracturing or even complete damage of the samples. \cite{16} For superconducting wires and tapes, in addition to structural damages, the huge stress can lead to irreversible degradation of the electromagnetic properties. \cite{19,20}
	The harm caused by stresses and strains seriously affects the use of superconducting devices. 
	As stated by Johansen, the failure of mechanical stability due to the stresses caused by pinning represents one of the greatest challenges in developing viable high-field high-temperature superconductor applications. \cite{16} 
	As a result, the magnetostriction behavior of IBSs has become an increasingly important issue. 
	
	In this work, we measured the magnetostriction of BaFe$_{1.908}$Ni$_{0.092}$As$_{2}$ single crystals down to 1.37 K with the magnetic field applied along different in-plane directions using pulsed magnetic fields up to 60 T. The magnetostriction in the direction of the twin boundaries shows significant hysteresis and huge magnetostriction (of the order of 10$ ^{-4} $). 
	The magnetization measurements for fields applied along the direction of the twin boundaries were carried out in static magnetic fields up to 33 T.
	The results evidence that a high critical-current density correlates with strong bulk pinning effects.   
	The magnetostriction can be qualitatively explained by using the magnetization data with an exponential model for the critical-current density. 
	As a result, the magnetostriction behavior can be well explained by the force internally exerted within the crystal due to flux pinning. 
	
	
	\begin{figure}[b]
		\includegraphics[width=8cm,height=12cm]{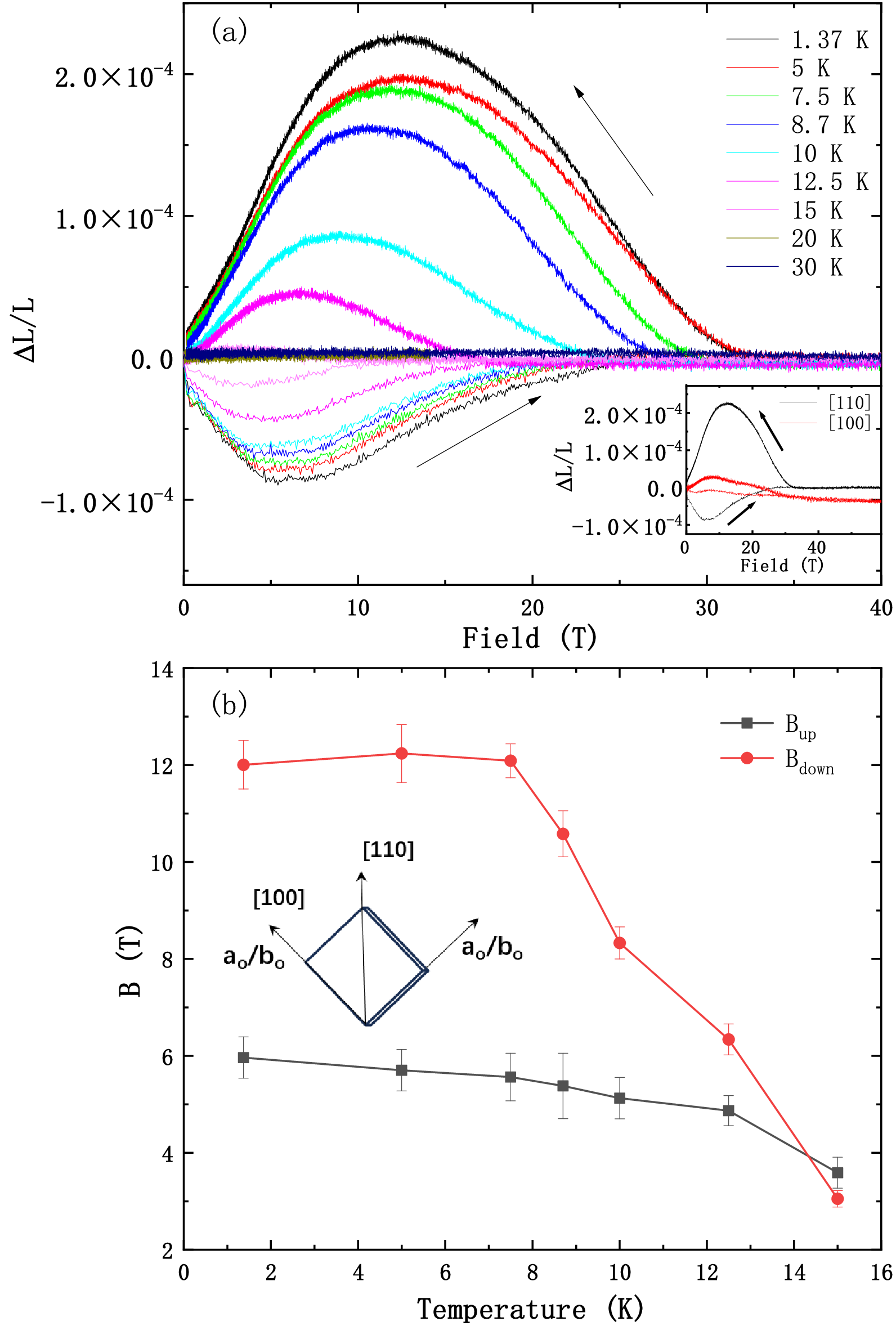}
		\caption{(a) Magnetic-field dependence of the longitudinal magnetostriction of BaFe$_{1.908}$Ni$_{0.092}$As$_{2}$ at different temperatures with magnetic field aligned along the orthorhombic [110] axis. The inset shows the magnetic-field dependence of the longitudinal magnetostriction for magnetic fields applied along the orthorhombic [110] and [100] axis at 1.37 K. (b) Temperature dependence of $ B_{up} $ and $ B_{down} $, with $ B_{up} $ and $ B_{down} $ denoting the magnetic fields where the maximum length deformation occurs for increasing and descending fields,  respectively. The inset shows the geometry of measurements.} \label{MS.eps}
	\end{figure}
	
	BaFe$_{1.908}$Ni$_{0.092}$As$_{2}$ single crystals were grown by the FeAs self-flux method, as described previously. \cite{05}
	The single crystals undergo a tetragonal-to-orthorhombic structural transition at a temperature $ T_{s} $ above the superconducting transition $ T_{c} $ ($\approx$ 19.3 K) ($T_{s}$ \textgreater \  $T_{c}$). \cite{21}  
	The orthorhombic distortion results in the formation of 45°-type twin domains, and the direction of the twin boundaries is fixed to the orthorhombic [110] direction. \cite{22}
	Large single crystals were cut along the orthorhombic [100] axis into a square with the size of $2 \times 2 \times 0.3$ mm$^{3}$ (length $\times$ width $\times$ thickness) as samples used for the magnetostriction and magnetization measurements. 
	The magnetostriction experiments were done using an optical fiber Bragg grating (FBG) method at the Dresden High Magnetic Field Laboratory. \cite{23,d}
	A 65-T nondestructive pulsed magnet driven by a capacitor bank with a pulse duration of about 150 ms was employed. \cite{24}
	The magnetization was measured using a homemade vibrating sample magnetometer in the High Magnetic Field Laboratory of Chinese Academy of Science. 
	A watercooled magnet which generates a maximum magnetic field of 35 T was used. 
	
	\begin{figure*}[htp]
		\includegraphics[height=5cm,width=17cm]{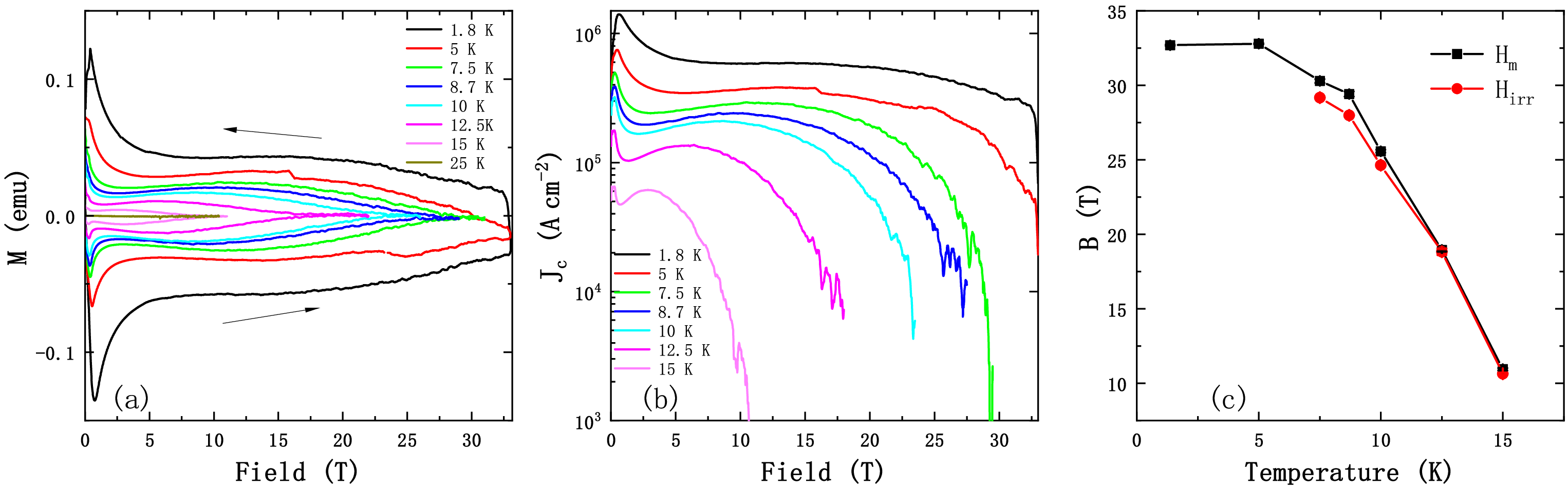}
		\caption{ (a) Magnetic-hysteresis loops of BaFe$_{1.908}$Ni$_{0.092}$As$_{2}$ at  different temperatures with magnetic fields up to 33 T applied along the orthorhombic [110] axis. (b) Field dependence of the calculated critical-current density based on the data shown in Fig.\ \ref{MH.eps}(a). (c) Temperature dependence of $ H_{m} $ and the irreversible field ($ H_{irr} $) extracted from the magnetostriction and magnetization measurements, respectively.} \label{MH.eps}
	\end{figure*}
	
	
	We measured the longitudinal magnetostriction along the orthorhombic [110] and [100] axis of one BaFe$_{1.908}$Ni$_{0.092}$As$_{2}$ sample in magnetic fields from 0 to 60 T at 1.37 K, as shown in the inset of Fig.\ \ref{MS.eps}(a).  
	A larger and better resolved magnetostriction signal occurs along the [110] direction, along the twin boundaries of the sample.
	Contrary, only a small magnetostriction occurs along the [100] direction. 
	For the [110] direction, not only the magnetostriction coefficient $ \Delta $$ L $/$ L $ is large, of the order of some $ 10^{-4} $, but also a clear hysteresis appears. 
	
	Magnetic-field-dependent magnetostriction measured between 1.37 and 30 K in pulsed magnetic fields up to 40 T applied along the direction of orthorhombic [110] axis are shown in Fig.\ \ref{MS.eps}(a). 
	It is obvious that the largest magnetostriction of about $2.3 \times 10^{-4}$ appears at 1.37 K and 12.5 T. 
	Between 1.37 and 8.7 K, $ \Delta $$ L $/$ L $ exceeds $10^{-4}$ at 5 - 20 T. 
    With increasing temperature, the values of the magnetostriction and magnitude of the hysteresis gradually decrease and completely disappear near $ T_{c} $. 
	When the magnetic field increases from zero up to 40 T, the sample is first compressed and reaches a minimum length, then returns to its original size. In the down sweep, the sample is stretched.
	The field where the magnetostriction becomes nonzero and evolves a hysteresis for the down sweeps is denoted by $ H_{m} $. 
	At all temperatures, the maximum tensile deformation is larger than the maximum compressive deformation. 
	The magnetic fields with the maximum length changes for up and down sweeps are denoted as $ B_{up} $ and $ B_{down} $, respectively. 
	The temperature dependence of $ B_{up} $ and $ B_{down} $ are shown in Fig.\ \ref{MS.eps}(b).
	The influence of temperature on $ B_{down} $ is larger than on $ B_{up} $.  
	The pinning resistance of the pinning center decreases significantly with increasing temperature. \cite{27} 

	In order to understand the nonmonotonous magnetostriction and its field-sweep-direction dependence, we measured the magnetization of BaFe$_{1.908}$Ni$_{0.092}$As$_{2}$. 
	In Fig.\ \ref{MH.eps}(a), we present typical magnetic-hysteresis loops of BaFe$_{1.908}$Ni$_{0.092}$As$_{2}$  at different temperatures with static magnetic fields up to 33 T applied along the orthorhombic [110] axis.   
	Apparently, 33 T is not sufficient for the sample to achieve saturation at temperatures below 5 K. 
	The symmetric curves and the absence of flux jumps suggest that bulk pinning rather than surface-barrier pinning dominates the magnetization process of the sample. 
	
	We calculated the critical-current density $ J_{c} $ based on the simplified formula $ J_{c} = 20 \Delta M/[d(1-d/3l)] $, where $\Delta M$ is the width of the magnetic hysteresis loop [Fig.\ \ref{MH.eps}(b)]. \cite{28} Here, $ l $ and $ d $ are the length and thickness of the sample ($ l $ \textgreater \  $ d $), respectively. 
	The calculated $ J_{c} $ exceeds $10^{5}$ A/cm$ ^{2} $ at 10 K and 17 T, demonstrating that our BaFe$_{1.908}$Ni$_{0.092}$As$_{2}$ sample has good current-carrying ability. 
	Figure \ref{MH.eps}(c) shows the temperature dependence of $ H_{m} $ and the irreversible magnetic field ($ H_{irr} $) extracted from the magnetostriction and magnetization measurements, respectively.  
	$H _{m} $ saturates with decreasing temperature, similar to the  in-plane upper critical field ($ H^{ab}_{c2} $) determined by magnetotransport measurements in pulsed magnetic fields. \cite{07}
	\begin{figure*}[htp]
	\begin{minipage}{0.8\textwidth}
	    \includegraphics[width=\textwidth,height=9cm]{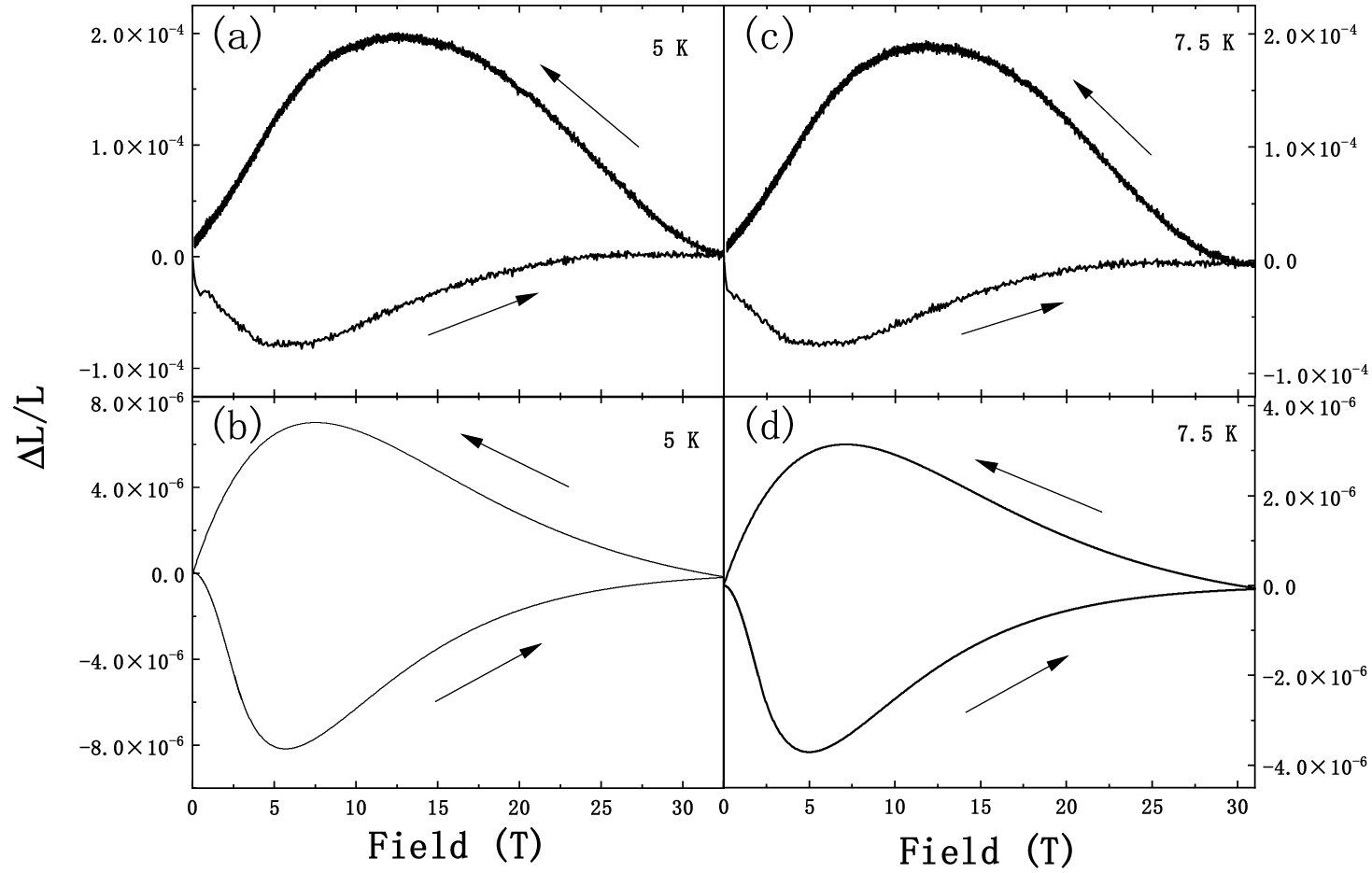}
    \end{minipage}%
    \begin{minipage}{0.2\textwidth}
    	\caption{Comparisons of measured and calculated $ \Delta $$ L $/$ L $ values on the basis of a pinning-induced mechanism at 5 K [(a) and (b)] and 7.5 K [(c) and (d)].}\label{Cal.eps}
    \end{minipage}
    \end{figure*}

	By comparing the experimental magnetostriction and magnetization data, the most reasonable explanation for the behavior of the magnetostriction is a flux-pinning-induced mechanism.
	Defects, impurities, or dislocations can form pinning centers. When the flux is pinned in the crystal, it exerts a force along the opposite direction in the sample. According to this mechanism, the sample is subject to a compressive force during the field up sweep and to an expansive force during the down sweep. In addition, since the flux motion is irreversible, the magnetostriction curve should be hysteretic. 
	With this in mind, the magnetostriction hysteresis loops can be estimated from the magnetization curves. 
	The basic equation that describes the sample-length change $ \Delta L $ under magnetic field is 
	
	\begin{eqnarray}
		\frac{\Delta L}{L}=-\frac{1}{2c\mu_0d}\int_{0}^{d}\left\{B_e^2-B^2(x)\right\}dx,
	\end{eqnarray}
	where $\mu_0$ is the permeability of vacuum, $ c $ is the elastic constant of the material along the $ x $ axis, $ B_{e} $ is the applied external magnetic field along the direction of $L$, $B(x)$ is the local magnetic-flux density and $L$ = $2d$ is the sample length. \cite{10}
	
	The presence of a local minimum in the magnetostriction curve during up sweep indicates that the pinning force ($ F_{P} $) has a maximum in the process. 
	In the Kim and Bean model, $ F_{P} $ increases monotonously with increasing external magnetic field. Only in the case when the critical-current density depends exponentially on magnetic field, $ F_{P} $ has a maximum and the magnetostriction can posses a minimum in the up sweep. \cite{11}
	Figure \ref{Cal.eps} shows the comparison of measured and calculated magnetostriction data based on the exponential model given in ref.\ [13] at 5 and 7.5 K (using the elastic constant $ c $ = 70 GPa). \cite{29}
	
	The calculated magnetostriction curves agree qualitatively well with the measured ones both for up and down sweeps. 
	The difference in the magnitude of the calculated and measured data is due to the fact that the exponential model ignores demagnetization effects and considers only one-dimensional flux penetration. \cite{10}
	The sample shape used and the direction of the field applied in our experiments are expected to produce strong demagnetization effects, which could make effectively applied magnetic field much bigger than applied external field and would cause larger magnetostriction signal. \cite{13}
	
	
	In summary, we measured the magnetostriction and the magnetization of single-crystalline BaFe$_{1.908}$Ni$_{0.092}$As$_{2}$ using pulsed magnetic fields up to 60 T and static magnetic fields up to 33 T, respectively.  
	We observe a huge longitudinal magnetostriction (of the order of 10$ ^{-4} $) when the field is aligned along the direction of the twin boundary. 
	The magnetization measurements suggest high critical-current densities with strong bulk pinning. 
	By using magnetization data with a model where the critical-current density depends exponentially on magnetic field \cite{11}, we can reproduce the magnetostriction curves qualitatively. 
	As a result, the magnetostriction can be well explained by the force exerted within the crystal due to the flux pinning. Such huge magnetostriction effect is a non-negligible issue when using 122-type compounds to manufacture magnets and devices for high-field applications.

	 This work was supported by the National Natural Science Foundation of China (Grants No. 11704385, No. 11874359), the National Key Research and Development Program of China (Grant No. 2018YFA0704200), the Strategic Priority Research Program (B) of the CAS (Grants Nos. XDB25000000), the Youth Innovation Promotion Association of CAS (Grant No. Y202001). We acknowledge the support of the HLD at HZDR, a member of the European Magnetic Field Laboratory (EMFL) and by the Deutsche Forschungsgemeinschaft (DFG) through the W{\"u}rzburg-Dresden Cluster of Excellence on Complexity and Topology in Quantum Matter ct.qmat (EXC 2147, Project No. 390858490). A portion of this work was performed on the Steady High Magnetic Field Facilities, High Magnetic Field Laboratory, Chinese Academy of Sciences, and supported by the High Magnetic Field Laboratory of Anhui Province.
	%

\end{document}